# Effect of exchange-correlation and core-electron approximations on the calculated superconducting transition temperature of palladium hydride

Samaneh Sadat Setayandeh[1], Tim Gould[1], Toktam Morshedloo[2,3] and Evan Gray[1,*]

[1]Queensland Micro- and Nanotechnology Centre, Griffith University, Brisbane, Australia

[2]Department of Physics, Payame Noor University, Tehran, Iran

[3]Nano Structured Coatings Institute of Yazd Payame Noor University, Yazd, Iran

**Abstract**

Realistic prediction of the superconducting transition temperature ($T_c$) for PdH is a long-standing challenge, because it depends on robust calculations of the electron and phonon band structures to obtain the electron-phonon scattering matrix element. To date, first-principles calculations employing density functional theory (DFT) have been based on selected exchange-correlation and core-electron approximations. Incorporating anharmonicity produced a more realistic value of $T_c$ that nevertheless still disagreed strongly with experiment unless adjustable parameters were introduced. Here we consider how the value of $T_c$ predicted using DFT in the harmonic approximation depends on the DFT scheme employed. The rationale for this work is that unless the calculation of $T_c$ within the harmonic approximation is robust, albeit incorrect, there is not a solid foundation for incorporating anharmonicity meaningfully. Six combinations of exchange-correlation approximation (LDA, PBE, PBEsol) and core-electron approximation (PAW, USPP) were tested. Following a carefully systematic methodology, the calculated $T_c$ was found to vary by a factor exceeding two across the tested DFT schemes. This suggests strongly that "standard" DFT, even including anharmonicity, is not reliable for PdH, implying that a higher-rung method will be needed to calculate a realistic lattice constant and phonon band structure, and so predict $T_c$ accurately.

___________

* Corresponding author. E-mail address: e.gray@griffith.edu.au

# 1. Introduction

Palladium hydride, PdH$_x$ where $x \approx 0.6$ is easily reached at room temperature and atmospheric pressure [1], is the ancestor of all interstitial metal hydrides and a model for understanding how this class of hydrogen storage materials works. Stoichiometric PdH is formed only at hydrogen pressures above about 1 GPa at room temperature [2-4], but is much easier to study by theory and computation owing to its perfect cubic symmetry.

For this reason, density functional theory (DFT) has often been employed to study the electron and phonon band structures of stoichiometric PdH, describing the electrons and ion cores by various types of approximations [5-19]. As noted in a recent review [20], the electronic properties are predicted with good agreement between the many approaches employed and reported. On the other hand, agreement among published phonon calculations, and between calculations and experiments, is so poor for the optical phonon bands [20], that only qualitative interpretations of the physical properties of PdH$_x$ known from experiments can be made. This is especially true of superconductivity, exhibited by PdH$_x$ for $x > 0.7$ approximately, which exhibits a reverse isotope effect between protium, deuterium and tritium [21, 22]: the most sophisticated modern first-principles calculation [16] estimates the superconducting transition temperature ($T_c$) of PdH at about half of the measured value.

Calculating $T_c$ for PdH thus persists as a significant challenge in condensed matter physics, even though outstanding successes have been reported recently in calculating the superconducting properties of high-temperature H-based phonon superconductors [23]. The basic problem derives from the need to calculate phonon properties from first principles in order to obtain the electron-phonon coupling constant ($\lambda$) and $T_c$, for which the Eliashberg-Migdal theory of strongly-coupled superconductors is usually employed in the Allen-Dynes formulation [24]. One fundamental problem, rarely acknowledged in the literature, is that within the harmonic approximation the DFT-derived phonon properties of PdH depend strongly on the DFT scheme employed [25], where "DFT scheme" indicates the combination of exchange-correlation approximation and core-electron approximation. Phonon properties then propagate into the Eliashberg spectral function (ESF), $\lambda$ and $T_c$ through the electron–phonon scattering matrix element and moments of the phonon frequency distribution. Another challenge is to treat anharmonicity of the H site potential, which was shown to be significant by neutron scattering experiments [26, 27] and appears to be the main cause of the disagreement between theory and experiment as to the value of $T_c$ [20].

Even more fundamentally, it was reported recently that the atomic volume of H in Pd, as calculated employing six common "standard" DFT schemes, disagreed very significantly with experiment [28], suggesting that even the total Born-Oppenheimer energy is not reliable, in consequence of which the phonon properties also cannot be relied on.

Here scrutiny is applied to calculating $T_c$ within the harmonic approximation, on the ground that incorporating anharmonicity into standard DFT schemes is fruitless if the results depend strongly on the chosen scheme. The authors' purpose was not to attempt a more realistic calculation of $T_c$, but to establish a baseline understanding of the variation in the values of $\lambda$ and $T_c$, obtained through straightforward application of the harmonic approximation, and varying according to the DFT scheme employed. The rationale was that, since both harmonic and anharmonic behaviours are influenced by the quality of the predicted potential energy surface, an analogous spread in calculated $T_c$ values might be expected when anharmonicity is introduced. A carefully systematic approach was followed, comparing the results obtained with various DFT schemes within the harmonic approximation, based on convergence tests, and employing the same value for the *ad-hoc* Coulomb pseudopotential ($\mu^*$). In contrast to some published studies, no adjustment was made to $\mu^*$ to force the calculations to match the experimentally measured $T_c$. Both octahedral (*oct*) and tetrahedral (*tet*) stoichiometric PdH were studied. The latter is interesting in this context because mathematically the potential at the smaller *tet* interstitial site can be expected to be much more closely harmonic than at the relatively large *oct* site.

## 2. Structures investigated

Two high-symmetry polymorphs of PdH were investigated: the rock-salt structure, PdH(*oct*), in which hydrogen (protium or its isotopes deuterium and tritium) occupies octahedral (*oct*) interstitial lattice sites (Fig. 1. (a)), and the zincblende structure PdH(*tet*) (Fig. 1. (b)), in which hydrogen occupies tetrahedral interstitial sites. Although experimental evidence points strongly to the *oct* structure being formed under commonly applied conditions of pressure and temperature, the total electronic energy of the *tet* structure calculated with DFT is typically slightly lower, with the order of stability reversed by adding zero-point energy of the nucleii [20]. Both structures are thus of interest from a conceptual point of view.

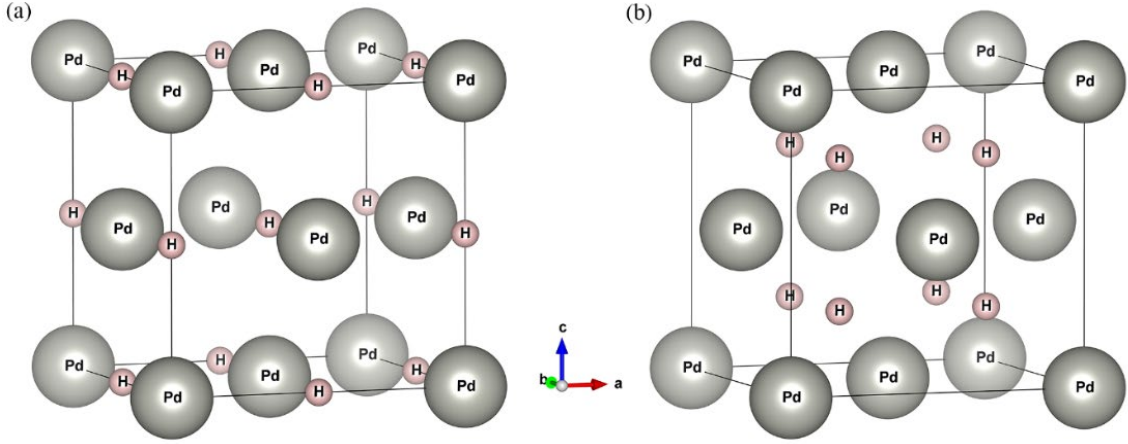

**Figure 1.** Interstitial sites in (a) PdH(*oct*) and (b) PdH(*tet*). Half occupancy of the *tet* sites corresponds to PdH(*tet*) and full occupancy would correspond to PdH$_2$.

## 3. Computational methods

The Quantum Espresso package [29] was used to perform *ab-initio* calculations for single unit cells of PdH, for various DFT schemes (combinations of density functional and core-electron approximations). Calculations were carried out with the local density approximation (LDA) [30], the generalised gradient approximation (GGA) by Perdew, Burke, and Ernzerhof (PBE) [31] and the PBEsol approximation [32, 33] for the exchange-correlation functional. In combination with the three exchange-correlation functionals, two popular core-electron approximations were applied: projector-augmented wave (PAW) [34] and ultra-soft pseudopotentials (USPP) [35]. A core-electron approximation is necessary to avoid difficult-to-treat rapid variations in the density and orbitals near the nucleus.

To sample the Brillouin zone, a Monkhorst-Pack mesh with a size of 12×12×12 was applied together with an energy cut-off of 40 Ry and a DEGAUSS value of 0.02, both selected after a series of careful convergence tests on the optimised structures. A sample convergence test to select the Monkhorst-Pack *k* mesh size for PdH(*oct*) is shown in Fig. 2. The DEGAUSS parameter specifies the degree of Gaussian smearing of the Fermi-Dirac function at $E_F$ at zero temperature, applied to overcome the discreteness of the *k*-point mesh [36]. The inset figure shows a sample convergence test to select a proper DEGAUSS value for PdH(*oct*). A value for DEGAUSS is seldom reported in the literature and DEGAUSS is commonly set to a "reasonable" value, such as 0.05. This was also the case in previous studies by this group, but explicitly testing its convergence led to a slightly lower value for DEGAUSS being used here, and slightly different results. For the phonon calculations, dynamical matrices [37] were calculated on a 4×4×4 *q*-point grid to obtain fast Fourier transformation of the real-space interatomic force-constants matrix, within the harmonic approximation.

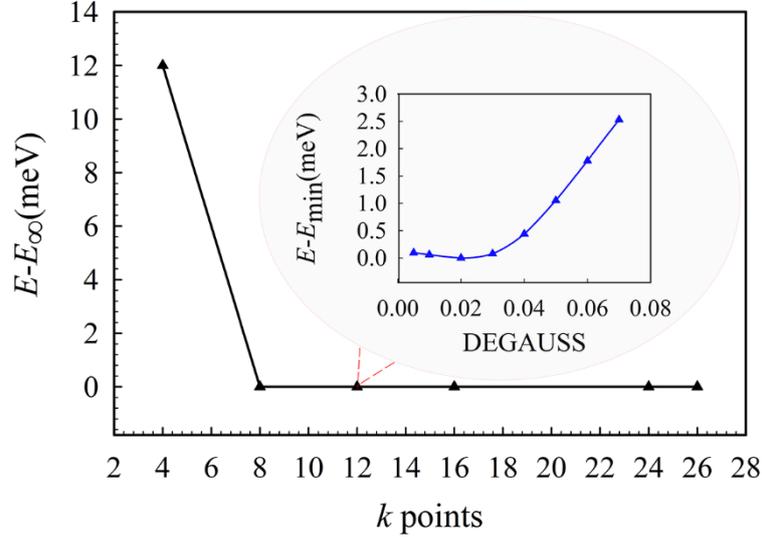

**Figure 2.** Convergence test for total energy with increasing size of Monkhorst-Pack mesh for PdH(*oct*) with the PBE/PAW scheme, relative to a 26-point mesh. A 12×12×12 mesh was selected for all subsequent calculations. The inset shows the convergence test for total energy when changing the value of DEGAUSS. DEGAUSS = 0.02 was employed for all subsequent calculations.

## 4. Results and discussion

### 4.1. Equilibrium lattice constant

Table I lists the equilibrium lattice constants of PdH (*oct*) and PdH (*tet*) polymorphs calculated with the six DFT schemes (combinations of exchange-correlation and core-electron approximation) obtained by minimising the total Born-Oppenheimer energy, employing the variable-cell optimization approach embedded in Quantum Espresso. A value of $1.0 \times 10^{-11}$ a.u. was used for the convergence threshold. The results follow expectations insofar as the equilibrium lattice constants calculated with LDA and GGA/PBE approximations are respectively lower and higher than experimental values, with the GGA/PBEsol approximation in between. The differences caused by choosing the PAW or USPP core-electron approximation were negligible.

**Table I.** Zero-temperature equilibrium lattice constants of PdH (oct) and PdH (tet) structures calculated with 6 different DFT schemes. For comparison, experimental lattice constants are 3.881 Å at 0 K for Pd [38] and 4.09 Å for PdH at 77 K [39].

| Structure | Exch-corr approx. | Core-electron approx. | $a$ (Å) |
|---|---|---|---|
| PdH(D/T)(o*ct*) | LDA (PZ) | PAW | 4.04 |
| | | USPP | 4.05 |
| | GGA (PBE) | PAW | 4.12 |
| | | USPP | 4.12 |
| | GGA (PBEsol) | PAW | 4.06 |
| | | USPP | 4.06 |
| PdH(D/T)(*tet*) | LDA (PZ) | PAW | 4.14 |
| | | USPP | 4.12 |
| | GGA (PBE) | PAW | 4.19 |
| | | USPP | 4.19 |
| | GGA (PBEsol) | PAW | 4.17 |
| | | USPP | 4.17 |

## 4.2. Energy barriers for H mobility

Figure 3 shows the relative energy for fixed geometry cells, with H placed at 0.25 (*tet*) and 0.33 (barrier) along the diagonal; relative to 0.5 (*oct*). This reveals that the barrier for H mobility varies significantly with choice of DFT scheme. In PBE the *oct* and *tet* sites have nearly degenerate electronic energies (this changes once nuclear energies are accounted for), whereas in LDA and PBEsol *tet* is nearly 50 meV higher. Moreover, in LDA the energies are sensitive to the choice of potential. We therefore see that forces (as represented by lattice parameters), and energies (shown directly in the figure) are quite sensitive to method choice. As we show in the next section, calculations of $T_c$ are even more sensitive.

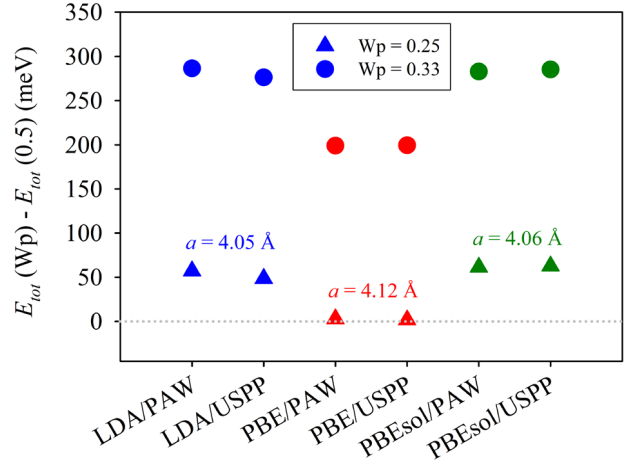

**Figure 3.** H energy barriers computed using different schemes. Wp is the Wyckoff position, measured along the diagonal of the conventional FCC unit cell, relative to the *oct* site (Wp = 0.5). Wp = 0.25 indicates the *tet* site and 0.33 is near the potential barrier maximum.

### 4.3. Superconducting transition temperature

Superconducting transition temperatures were calculated using the Allen-Dynes formula [24]:

$$T_c = \frac{f_1 f_2 \omega_{\log}}{1.2} \exp\left[\frac{-1.04(1+\lambda)}{\lambda - \mu^*(1+0.62\lambda)}\right] \quad (1)$$

where

$$\lambda = 2\int_0^\infty \alpha^2 F(\omega) \omega^{-1} d\omega \quad (2)$$

Here, $\mu^*$ and $\lambda$ are unit-less constants and $\alpha^2 F(\omega)$ is the ESF at angular frequency omega. Other terms will be defined below. In the literature various theoretical values have been calculated or assumed for $\lambda$ and $\mu^*$ in order to calculate $T_c$, sometimes to make the calculated value agree with experiment, as listed in Table II [16, 18, 19, 40-43].

**Table II.** Values reported for electron-phonon coupling constant ($\lambda$) and screened Coulomb potential ($\mu^*$) together with calculated superconducting transition temperatures ($T_c$). H/A indicates that the harmonic or an anharmonic approximation was used.

| Parameters | $\mu^*$ | $\lambda$ | $T_c$ (K) | Approximations; exch-corr/cor e$^-$ | Ref. |
|---|---|---|---|---|---|
| PdH(*oct*) | 0.1 | 0.560 | 9.5 | H; X$_\alpha$/APW | [42] |
|  | 0.085 | 0.560 | 8.5 | H; X$_\alpha$/APW | [42] |
|  | 0.085 | 1.55 | 47 | H; LDA/USPP | [16] |
|  | 0.085 | 0.40 | 5 | A; LDA/USPP | [16] |
|  | 0.18 | 0.8 | 9 | H; GGA/PAW | [19] |
|  | 0.085 | 0.449 | 5.2 | H; –/– | [41] |
|  | 0.085 | 0.54 | 7.9 | H; X$_\alpha$/APW | [11] |
|  | 0.085 | 0.500 | 8.0 | A; X$_\alpha$/APW | [43] |
| PdD(*oct*) | 0.1 | 0.599 | 9.2 | H; X$_\alpha$/APW | [42] |
|  | 0.085 | 0.599 | 9.8 | H; X$_\alpha$/APW | [42] |
|  | 0.085 | 1.55 | 34 | H; LDA/USPP | [16] |
|  | 0.085 | 0.46 | 6.5 | A; LDA/USPP | [16] |
|  | 0.085 | 0.62 | 9.8 | H; X$_\alpha$/APW | [11] |
|  | 0.085 | 0.603 | 10.5 | A; X$_\alpha$/APW | [43] |
| PdT(*oct*) | 0.1 | 0.628 | 8.7 | H; X$_\alpha$/APW | [42] |
|  | 0.085 | 0.628 | 10.2 | H; X$_\alpha$/APW | [42] |
|  | 0.085 | 1.55 | 30 | H; LDA/USPP | [16] |
|  | 0.085 | 0.48 | 6.9 | A; LDA/USPP | [16] |
|  | 0.085 | 0.678 | 11.8 | A; X$_\alpha$/APW | [43] |
| PdH(*tet*) | 0.062 | – | 8.82 | H; LDA/– | [18] |
| PdD(*tet*) | 0.062 | – | 11.05 | H; LDA/– | [18] |

In the present work a value of $\mu^* = 0.085$ was used for all the calculations, as calculated by Klein *et al.* [9] for PdH. Figure 4 illustrates the relatively minor negative effect of increasing $\mu^*$ on the argument $\{-1.04(1+\lambda)/[\lambda - \mu^*(1+0.62\lambda)]\}$ of the exponential in Eq. (1). It should be noted, however, that the lower phonon frequencies for D and T should lead to lower values for $\mu^*$ [24] and higher $T_c$ values. Unusually low values of 0.062 for H and 0.022 for D were needed to fit the harmonic calculation of $T_c$ to experiment in Ref. [18] in order to ascribe the inverse isotope effect purely to retardation. Ostanin *et al.* [19] used an unusually large value of 0.18 without explanation.

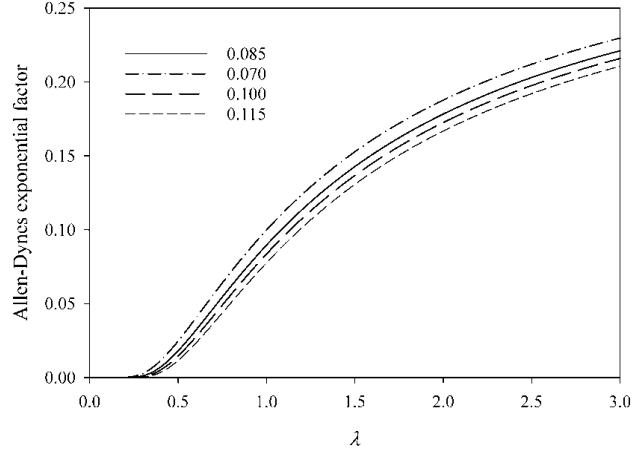

**Figure 4.** Effect of changing $\mu^*$ (values indicated) on the argument $\{-1.04(1+\lambda)/[\lambda-\mu^*(1+0.62\lambda)]\}$ of the exponential in Eq. (1).

The parameters $f_1$ and $f_2$ were calculated from [24]:

$$f_1 = \left[1+\left(\frac{\lambda}{2.46+9.35\mu^*}\right)^{3/2}\right]^{1/3} \quad (3)$$

$$f_2 = 1+\frac{\left[(\bar{\omega}_2/\omega_{\log})-1\right]\lambda^2}{\lambda^2+\left(1.82+11.5\mu^*\right)^2\left(\bar{\omega}_2^2/\omega_{\log}^2\right)} \quad (4)$$

where

$$\bar{\omega}_2^2 = \int_0^\infty \alpha^2 F(\omega)\omega\,d\omega \Big/ \int_0^\infty \alpha^2 F(\omega)\omega^{-1}d\omega \quad (5)$$

and

$$\omega_{\log} = \exp\left[\int_0^\infty \alpha^2 F(\omega)\omega^{-1}\ln\omega\,d\omega \Big/ \int_0^\infty \alpha^2 F(\omega)\omega^{-1}d\omega\right] \quad (6)$$

The electron-phonon coupling constant was calculated using the ESF produced by Quantum Espresso.

Calculations using these formulae were carried out for PdH/D/T(*oct*) and PdH/D/T(*tet*) using six different DFT schemes: LDA/USPP, PBE/USPP, PBEsol/USPP, LDA/PAW, PBE/PAW, PBEsol/PAW. Values for the calculated parameters are listed in Tables III and IV. The isotopic hydrides are also listed for completeness, since for a superconductor with $\mu^* \neq 0$, $T_c$ does not scale exactly as $M^{-1/2}$ in the harmonic approximation. Figure 5 illustrates graphically the wide variation in $T_c$ between DFT schemes.

**Table III.** Comparison of calculated values of $\omega_{\log}$, $\lambda$, $f_1f_2$ and $T_c$ for PdH(*oct*), PdD(*oct*) and PdT(*oct*) with six different DFT schemes.

| Structure | Exch-corr approx. | Core-electron approx. | $\omega_{\log}$ (THz) | $\lambda$ | $f_1f_2$ | $T_c$ (K) |
|---|---|---|---|---|---|---|
| PdH(*oct*) | LDA (PZ) | PAW | 7.88 | 1.018 | 1.060 | 30.7 |
| | | USPP | 9.43 | 0.816 | 1.048 | 25.6 |
| | GGA (PBE) | PAW | 4.10 | 2.584 | 1.343 | 45.5 |
| | | USPP | 5.72 | 1.814 | 1.178 | 44.9 |
| | GGA (PBEsol) | PAW | 9.85 | 0.711 | 1.039 | 19.9 |
| | | USPP | 7.22 | 0.862 | 1.048 | 21.4 |
| PdD(*oct*) | LDA (PZ) | PAW | 6.82 | 0.867 | 1.049 | 20.5 |
| | | USPP | 7.16 | 0.813 | 1.045 | 19.1 |
| | GGA (PBE) | PAW | 3.24 | 2.537 | 1.311 | 34.7 |
| | | USPP | 4.32 | 1.807 | 1.167 | 33.5 |
| | GGA (PBEsol) | PAW | 7.53 | 0.717 | 1.037 | 15.5 |
| | | USPP | 7.57 | 0.710 | 1.036 | 15.2 |
| PdT(*oct*) | LDA (PZ) | PAW | 6.27 | 0.775 | 1.040 | 15.2 |
| | | USPP | 6.10 | 0.820 | 1.044 | 16.5 |
| | GGA (PBE) | PAW | 2.76 | 2.530 | 1.301 | 29.3 |
| | | USPP | 3.64 | 1.821 | 1.164 | 28.3 |
| | GGA (PBEsol) | PAW | 6.43 | 0.712 | 1.035 | 13.0 |
| | | USPP | 6.46 | 0.710 | 1.035 | 13.0 |

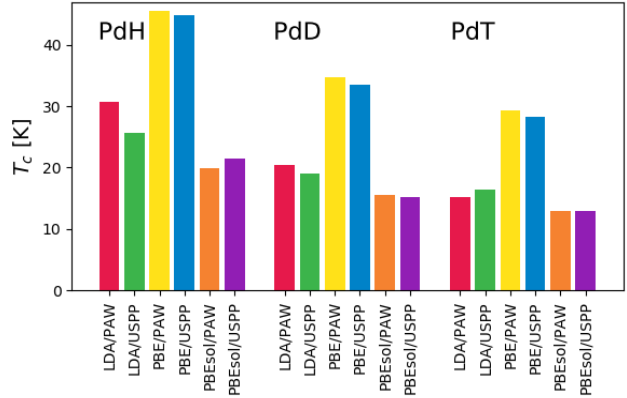

**Figure 5.** The $T_c$ values for PdH/D/T(*oct*) calculated with six different DFT schemes.

Referring first to the controlled and systematic survey reported here, in the *oct* case, the calculated superconducting transition temperatures were very sensitive to the effect on the phonon frequency distribution coming from the different lattice parameters obtained by minimising the total Born-Oppenheimer energy according to different exchange-correlation approximations, but fairly similar between the two core-electron approximations tested, albeit with a greater influence on $T_c$ than on the lattice constants.

The general sensitivity problem is further illustrated by the comparison between the LDA/USPP value for PdH(*oct*) in Table III ($T_c$ = 26 K) and that reported by Errea et. al. [16] for the same DFT scheme employing the harmonic approximation ($T_c$ = 47 K). All the calculated $T_c$ values are much higher than the experimental ones for PdH (8–9 K) [21, 44]. Errea *et al.* [16] suggested that "soft" H modes, in which the harmonic potential felt by the H is small relative to higher-order (*e.g.* cubic or quartic) terms, led to their large values of $\lambda$ and $T_c$. Soft modes can be detected by low or imaginary frequencies in the phonon spectra, or by strong sensitivity to small perturbations of the atom. Our results (the full phonon dispersion curves from a previous series of calculations are displayed in Ref. [25]) showed no imaginary frequencies, most likely because the equilibrium zero-temperature lattice constant was smaller and thus led to "harder" confinement of the H.

In Table III the PBE/PAW and PBE/USPP values for $T_c$ are anomalously high compared to the other DFT schemes. This reflects the qualitatively different phonon curves (see top panel of Figure 7) found using PBE *vs* LDA/PBEsol, especially the lack of a sharp peak at around 10 THz. To put the magnitude of this difference in context, as reported many times since the 1970s, the isotope effect as calculated in the harmonic approximation has the wrong sign compared to experiment, which is the main indication that anharmonicity is important. As shown clearly by Fig. 5, the effect of the DFT scheme on $T_c$ is greater than the isotope effect.

In the Allen-Dynes formulation $T_c$ is proportional to $f_1f_2$. Yussouff et. al [43] proposed that the approximation $f_1f_2 \approx 1.03$ was valid to within 10%, and this is largely borne out by Tables III and IV where the full Allen-Dynes values of $f_1f_2$ for *oct* occupancy were mostly slightly larger and those for *tet* occupancy were mostly slightly smaller than 1.03. The exception was the PBE approximation, where high values of $\lambda$ led to the highest values of $f_1f_2$ and $T_c$. The values for $f_1f_2$ are summarised in Fig. 5.

Turning now to the *tet* polymorph, smaller values of $\lambda$ and $T_c$ were obtained (Table IV) with the same value for $\mu^*$ and there was less variation between DFT schemes. This may be understood by referring to plots of the ESF and $\lambda(\omega)$ in Fig. 6 below, where a much stronger bias to high-frequency modes and a much smaller variation between DFT schemes are seen.

Figure 7 shows the phonon spectra, $\lambda(\omega) = 2\int_0^\omega \alpha^2 F(\omega')\omega'^{-1}d\omega'$ and the ESF for PdH(*oct*) (top) and PdH(*tet*) (bottom) with the six possible DFT schemes. It is apparent that the high $\lambda$ and $T_c$ values obtained with the PBE approximation come from the (fictitious) absence of a phonon band gap. The rapid increase in $\lambda(\omega)$ at medium frequencies (< 10 THz) leads to relatively high $T_c$ values because the damping effect of $\omega^{-1}$ in Eqs (2) and (6) is lessened.

The same argument (the damping effect of $\omega^{-1}$) explains both the lower values of $\lambda$ and $T_c$ predicted by all DFT schemes for the *tet* structure and the diminished variation between schemes: the much higher optical phonon frequencies contribute less to both $\omega_{\log}$ and $\lambda$ relative to the *oct* case and differences between them matter less.

**Table IV.** Comparison of calculated values of $\omega_{\log}$, $\lambda$, $f_1f_2$ and $T_c$ for PdH(*tet*), PdD(*tet*) and PdT(*tet*) with six different DFT schemes.

| Structure | Exch-corr approx. | Core-electron approx. | $\omega_{\log}$ (THz) | $\lambda$ | $f_1f_2$ | $T_c$ (K) |
|---|---|---|---|---|---|---|
| PdH(*tet*) | LDA (PZ) | PAW | 3.91 | 0.587 | 1.031 | 4.9 |
| | | USPP | 3.84 | 0.605 | 1.032 | 5.2 |
| | GGA (PBE) | PAW | 4.51 | 0.519 | 1.026 | 3.8 |
| | | USPP | 4.45 | 0.548 | 1.028 | 4.5 |
| | GGA (PBEsol) | PAW | 4.10 | 0.549 | 1.028 | 4.2 |
| | | USPP | 3.87 | 0.587 | 1.031 | 4.8 |
| PdD(*tet*) | LDA (PZ) | PAW | 3.53 | 0.582 | 1.032 | 4.3 |
| | | USPP | 3.55 | 0.606 | 1.034 | 4.8 |
| | GGA (PBE) | PAW | 4.12 | 0.524 | 1.027 | 3.6 |
| | | USPP | 3.99 | 0.547 | 1.029 | 4.0 |
| | GGA (PBEsol) | PAW | 3.60 | 0.555 | 1.030 | 3.8 |
| | | USPP | 3.40 | 0.592 | 1.033 | 4.3 |
| PdT(*tet*) | LDA (PZ) | PAW | 3.35 | 0.583 | 1.032 | 4.1 |
| | | USPP | 3.30 | 0.599 | 1.034 | 4.3 |
| | GGA (PBE) | PAW | 3.89 | 0.526 | 1.027 | 3.4 |
| | | USPP | 3.72 | 0.544 | 1.029 | 3.7 |
| | GGA (PBEsol) | PAW | 3.72 | 0.547 | 1.029 | 3.7 |
| | | USPP | 3.45 | 0.577 | 1.031 | 4.1 |

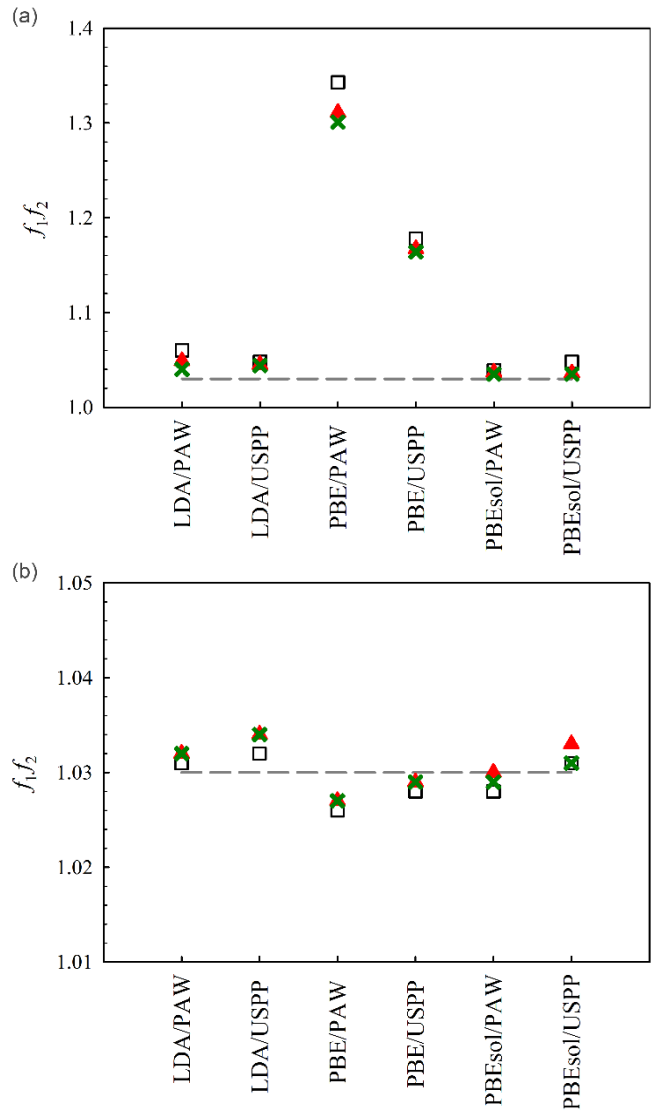

**Figure 6.** Comparison of $f_1f_2$ values from six different DFT schemes for (a) PdH/D/T(*oct*) and (b) PdH/D/T(*tet*). Open squares: PdH; filled triangles: PdD; crosses: PdT. The dashed line represents the value of 1.03 estimated by Yussouff *et al.* [43].

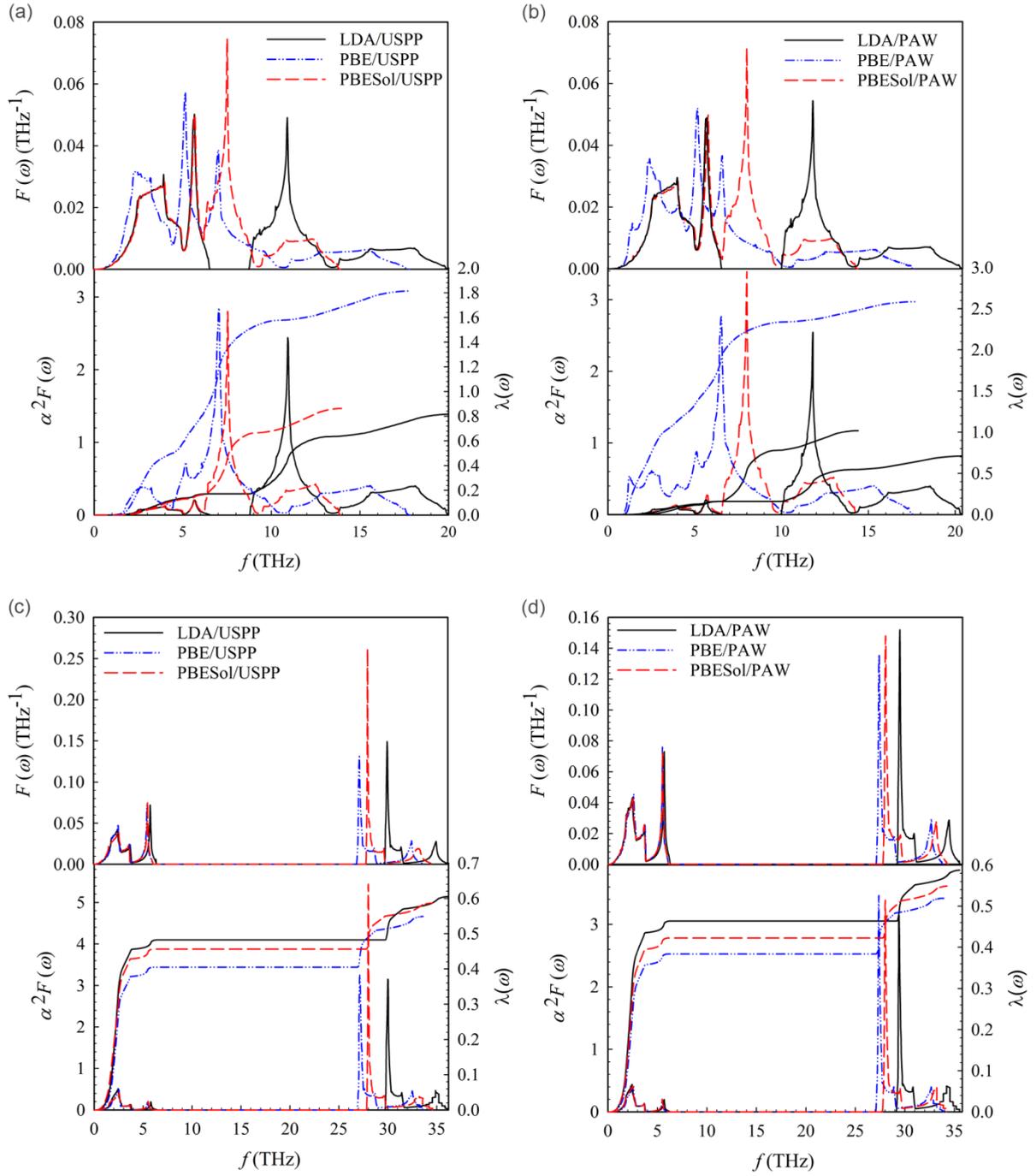

**Figure 7.** Upper panels: phonon DoS; lower panels: ESF (solid lines) and *λ(ω)* (dashed lines) of (a) PdH(*oct*) obtained with USPP, (b) PdH(*oct*) obtained with PAW, (c) PdH(*tet*) obtained with USPP and (d) PdH(*tet*) obtained with PAW and three different exchange-correlation approximations. The phonon DoS and ESF are shown in Hertzian rather than radian frequency units for ease of appreciation.

## 5. Summary and conclusions

A carefully systematic comparison was made between *ab-initio* calculations of the superconducting transition temperature ($T_c$) of PdH/D/T, assuming that hydrogen (protium, deuterium or tritium) occupies octahedral (*oct*) or tetrahedral (*tet*) interstitial lattice sites. Six popular DFT schemes, combining the LDA, PBE and PBEsol exchange-correlation approximations with USPP and PAW core-electron approximations, were tested within the harmonic approximation. The same value for the Coulomb pseudopotential was used to describe retardation in all cases, to expose the differences arising solely in the details of the DFT scheme.

In the *oct* case, wide disagreements were found between the electron–phonon coupling constants ($\lambda$) and average phonon frequencies ($\omega_{\log}$) calculated with different DFT schemes within the harmonic approximation, confirming the sensitivity of the calculated $T_c$ to the different lattice constants obtained by minimising the total Born-Oppenheimer energy at zero temperature. Changing the exchange-correlation approximation had a greater effect than changing the core-electron approximation, however both introduced significant variation to $T_c$. All schemes overestimated the value of $T_c$ relative to experiment in the *oct* case.

In the *tet* case, the optical phonon frequencies were pushed high enough that they contributed less (relative to *oct*) to the values $\lambda$, $\omega_{\log}$ and $T_c$, as well as muting the differences between DFT schemes.

In the *oct* case the spread of $T_c$ values between DFT schemes exceeded 2 to 1 for the same isotope: *e.g.* for PdH 45.5 K for PBE/PAW *versus* 19.9 K for PBEsol/PAW. In comparison, the values obtained by Errea *et al.* [16] after accounting for anharmonicity within one DFT scheme (LDA/USPP, predicting 5.0 K for PdH and 6.5 K for PdD) differ by a smaller factor from the accepted experimental values (around 10 K), raising the question of what results would be obtained with the other DFT schemes tested here. The need to incorporate anharmonicity has been demonstrated emphatically by Errea *et al.*, but the absolute values produced for $T_c$ are evidently not meaningful because a different DFT approach would produce different values.

This study has thus exposed a serious consequence of the basic problem of "standard" DFT applied to PdH uncovered by Setayandeh *et al.* [28], which is that not only are absolute lattice constants variably predicted between DFT schemes, the predicted atomic volume of H in Pd is in strong disagreement with experiment. Because phonon frequencies depend strongly on the absolute lattice constant, all predicted phonon-related properties must also vary depending on the particular DFT scheme employed. Since these different predictions cannot all be correct, some, perhaps all, are wrong in the sense that they do not represent the true outcome of applying the harmonic approximation. Accounting for anharmonicity on this basis therefore cannot be expected to succeed in calculating an accurate value for $T_c$ except by chance.

This fundamental problem deserves further work to try to find a DFT-based approach that yields accurate lattice constants as a first step, to provide a foundation for the subsequent incorporation of anharmonicity, then hopefully at last yield phonon dispersion curves and a phonon energy gap that agree in detail with inelastic neutron scattering results. The ultimate test of such a new approach would be to predict $T_c$ for PdH accurately from first principles, without fitted parameters.

**Acknowledgement**
SSS acknowledges receipt of a Griffith University PhD scholarship and Publication Assistance Scholarship.